\documentclass[11pt]{article}%
\usepackage{amssymb}
\usepackage{amsmath}
\usepackage{graphics}%
\usepackage{psfrag}
\usepackage{epsfig}
\usepackage{citesort}
\def\preprint%
#1{\thispagestyle{empty}~\newline\vspace*{-22.65mm}%
\begin{flushright}%
\begin{tabular}{l} #1%
\end{tabular}%
\end{flushright}%
\vspace{1cm}}%
\begin{document}%
\markright{\hfil Non-Locality, Contextuality and Transition Sets}%
\title{\bf \LARGE Non-Locality, Contextuality and Transition Sets}%
\author{Hans Westman\thanks{\tt Email: hwestman@perimeterinstitute.ca}\\[2mm]%
{\small \it Perimeter Institute for Theoretical Physics}\\%
{\small \it 31 Caroline Street North, N2L 2Y5 Waterloo, Ontario, Canada}\\[2mm]%
}%
\date{{\small \today }}%
\maketitle%
\begin{abstract}%
We discuss quantum non-locality and contextuality using the notion of transition sets \cite{Valentini,Valentinib}. This approach provides a way to obtain a direct logical contradiction with locality/non-contextuality in the EPRB gedanken experiment as well as a clear graphical illustration of what violations of Bell inequalities quantify. In particular, we show graphically how these violations are related to measures of non-local transition sets. We also introduce a new form of contextuality, {\em measurement ordering contextuality}, i.e. there exists commuting operators $\hat{\mathcal{A}}$ and $\hat{\mathcal{B}}$ such that the outcome for $\hat{\mathcal{A}}$ depends on whether we measured $\hat{\mathcal{B}}$ before or after  $\hat{\mathcal{A}}$. It is shown (excluding retro-causal and/or conspiratorial theories)  that any hidden variable theory capable of reproducing the quantum statistics has to have this property. This generalizes yet another feature of the hidden variable theory of deBroglie and Bohm.\\\\%
\end{abstract}%
\def\g{\mbox{\sl g}}%
\def\x{\mbox{\boldmath $x$}}%
\def\p{\mbox{\boldmath $p$}}%
\def\w{\mbox{\boldmath $w$}}%
\def\0{\mbox{\boldmath $0$}}%
\def\u{\mbox{\boldmath $u$}}%
\def\R{\rm I\!R}%
\def\bnabla{\mbox{\boldmath $\nabla$}}%

\section{Introduction}
Quantum non-locality has been, and still is, subject to lively debate. Some researchers tend to take the view that {\it `To those for whom non-locality is anathema, Bell's Theorem finally spells the death to the hidden-variable program'} \cite{Mermin}, while Bell himself was of a quite different opinion. He thought this conclusion was premature because even standard quantum mechanics is non-local in the sense that it fails to satisfy a mathematically precise and physically reasonable definition of local causality (see e.g. \cite[pp. 55]{Bell} or \cite{Cuisine}).

Quantum contextuality took on a livelier debate among physicists especially after simple proofs were constructed \cite{Peres,Mermin}. It is believed by some that contextuality cannot be a natural property of a hidden variable theory \cite{Mermin,Peres}. But also here Bell \cite[pp. 8--9]{Bell} was of a different view. Bell had a working knowledge of the pilot-wave theory of deBroglie and Bohm \cite{deBroglie,Crossroads,Bohm} in which contextuality is quite banal. In that theory quantum measurements are not passive interventions but disturb the system and play and active part in forming the outcome. This led Bell to criticize the use of the word `measurement' \cite{Measurement} rather than negating the possibility of hidden variable theories. Bell's train of thought is easily understood when contextuality is examined using the concrete hidden variable model of deBroglie and Bohm. In a forthcoming companion paper we will provide an extensive  exposition of quantum contextuality in deBroglie-Bohm theory.

In this article we will use the notion of transition sets \cite{Valentini,Valentinib} to discuss quantum non-locality and contextuality for arbitrary deterministic hidden variable theories. Indeterministic theories such as Nelson's stochastic mechanics \cite{Nelson} will not be considered here. This article is organized as follows. In section \ref{Trans} we define the notion of transition sets in the EPRB gedanken experiment \cite{EPRB}. Together with a definition of non-contextuality given in section \ref{CvsNL} we exhibit a direct logical contradiction with  non-contextuality/locality in section \ref{logical}. In section \ref{relation} we clarify what we mean by `violations of Bell inequalities' and proceed in section \ref{Quantify} to graphically illustrate how violations of Bell inequalities provide a lower bound on certain regions of non-local transition sets measures. In section \ref{communication} we reproduce a result by Pironio \cite{Pironio} demonstrating that violations of Bell inequalities yield a lower bound on the amount of classical communication required to reproduce the quantum statistics in the EPRB gedanken experiment. In section \ref{MOC} we show that any hidden variable theory (excluding retro-causal and/or conspiratorial theories) must be measurement ordering contextual, i.e. there exists commuting operators $\hat{\mathcal{A}}$ and $\hat{\mathcal{B}}$ such the the outcome of $\hat{\mathcal{A}}$ depends whether $\hat{\mathcal{B}}$ was measured first or after. We also provide a graphical illustration of the signal-locality theorem by Valentini \cite{Valentini} in section \ref{signalocality}.
\section{Transition sets in the EPRB gedanken experiment}\label{Trans}
\begin{figure}
\begin{center}
\psfrag{Lambda space}{{\Large $\Lambda$-space}}
\epsfig{figure=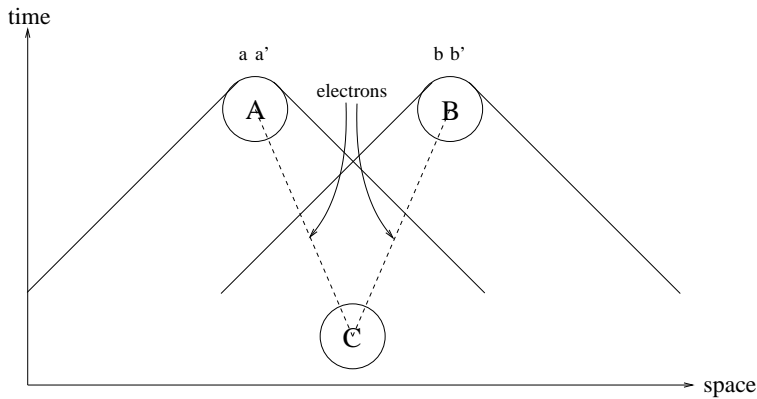,width=10cm}
\end{center}
\caption{Spacetime diagram of the EPR-Bohm gedanken experiment.}
\label{EPRB}
\end{figure}

Let us proceed to the notion of transition sets introduced in \cite{Valentini}. For simplicity, we confine the discussion to the EPRB scenario \cite{EPRB} (see fig. \ref{EPRB} for illustration). Consider two electrons prepared in a singlet state at a spacetime region ${\sf C}$. The two electrons are assumed to pass through two spatially separated spacetime regions ${\sf A}$ and ${\sf B}$ each containing a Stern-Gerlach apparatus. At ${\sf A}$ we can choose to either align the apparatus along the direction ${\bf a}$ or ${\bf a'}$ and at ${\sf B}$, ${\bf b}$ or ${\bf b'}$. We will confine ourself to situations for which ${\bf a}$, ${\bf a'}$, ${\bf b}$, and ${\bf b'}$ are perpendicular to the axis joining the two apparatuses. Therefore only an angle $a$, $a'$, $b$, $b'\in[0,2\pi[$ is sufficient to specify a direction ${\bf a}$, ${\bf a'}$, ${\bf b}$, ${\bf b'}$.

Let $\Lambda$ be the space of possible hidden variable\footnote{The terms `hidden variable' and `beable' are taken to be synonymous throughout this article. Therefore there is nothing `hidden' about these variables. They are the things which table and chairs are made of \cite[ch. 7]{Bell}.} configurations $\lambda\in\Lambda$  and let $A=\pm1$ and $B=\pm1$ represent the outcomes at region ${\sf A}$ and ${\sf B}$ respectively. In a deterministic hidden variable theory these outcomes $A$ and $B$ are not random but determined by some functions of the angles $a$ and $b$ and the variable $\lambda$. The probabilities for the outcomes $A=\pm1$ and $B=\pm1$ can then in principle be calculated once a probability distribution $\rho(\lambda)$ on $\Lambda$ has been given. Perhaps the most natural form of the functions are $A=A(a,\lambda)$ and $B=B(b,\lambda)$. But if we assume that the distribution is independent of the angles $\rho(\lambda;a,b)=\rho(\lambda)$ (i.e. $a$ and $b$ are `free variables' \cite{Cuisine,Shimony} so that conspiratorial and retro-causal models are excluded) then the Bell inequalities are satisfied \cite{Bell,Cuisine,Shimony}. Thus hidden variable theories of this type cannot cannot reproduce the quantum statistics and we are forced to either drop the assumption of free variables and/or consider more general functions $A(a,b,\lambda)$ and $B(a,b,\lambda)$ for which the outcome at one wing depends non-locally on the setting of the apparatus at the other. In this article we will keep the assumption of free variables $\rho(\lambda;a,b)=\rho(\lambda)$ and and let the functions be of the non-local form  $A(a,b,\lambda)$ and $B(a,b,\lambda)$. This type of non-local hidden variable theories can reproduce the quantum statistics.

 However, the form of the functions $A(a,b,\lambda)$ and $B(a,b,\lambda)$ is not the most general one. A more general form is $A([a,\mu],[b,\nu],\lambda)$ and $B([a,\mu],[b,\nu],\lambda)$, where $\mu$ and $\nu$ are additional variables (beables) perhaps belonging to the apparatuses at ${\sf A}$ and ${\sf B}$. The extra variables, which need not be statistically correlated with the $\lambda$'s, could represent variables that are hard or impossible to control for experimentalists. They could also represent macroscopic variables which can easily be controlled. For example, in the simple deBroglie-Bohm model considered in \cite{Valentini} the outcomes can also depend on the coupling strength $g_A$ and $g_B$ of the measurement apparatuses at ${\sf A}$ and ${\sf B}$. The functions therefore take the form $A([a,g_A],[b,g_B],\lambda)$ and $B([a,g_A],[b,g_B],\lambda)$ in that model. In the case where the variables $\mu$ and $\nu$ cannot easily be controlled one has to provide probability distributions $\rho_1(\mu)$ and $\rho_2(\nu)$. It is important to realize that these distributions are not necessarily independent of the angles. Since these variables might belong to the apparatuses the distributions could be directly influenced by the apparatus settings, i.e. we have $\rho_1=\rho_1(\mu;a)$ and $\rho_2=\rho_2(\nu;b)$. In distinction to the case where the probability distribution over $\Lambda$ depends on the angles, $\rho(\lambda)=\rho(\lambda;a,b)$, any angular dependence of the distributions $\rho_1(\mu;a)$ and $\rho_2(\nu;b)$ would not amount to any `conspiracy' nor indicate a retro-causal influence. Therefore, a possible angular dependence of the distributions $\rho_1$ and $\rho_2$ is not excluded by any assumption already made. In this article we shall for the sake of simplicity assume the simpler form $A(a,b,\lambda)$ and $B(a,b,\lambda)$. A more general analysis can presumably be carried out.

Let us therefore proceed to define a non-local transition set $T^{a\leftrightarrow a'}_{\ \ b}\subset\Lambda$ by 
\begin{eqnarray}\label{transdef}
T^{a\leftrightarrow a'}_{\ \ b}=\left\{\lambda\in\Lambda|B(a,b,\lambda)\neq B(a',b,\lambda)\right\}. 
\end{eqnarray}
If $\lambda\in T^{a\leftrightarrow a'}_{\ \ b}$ then the outcome at B will depend non-locally on the choice of angle ($a$ or $a'$) at A. Similarly, the transition set
\begin{eqnarray}
T^{\ a}_{b\leftrightarrow b'}=\left\{\lambda\in\Lambda|A(a,b,\lambda)\neq
A(a,b',\lambda)\right\}
\end{eqnarray}
means that if $\lambda\in T^{\ a}_{b\leftrightarrow b'}$ then the outcome at A will depend non-locally on the choice of angle at B. In a similar manner we define the transition sets $T^{a\leftrightarrow a'}_{\ \ b'}$ and $T^{\ a'}_{b\leftrightarrow b'}$. Notice the convention of putting the $a$-angles upstairs and the $b$-angles downstairs. Of course, in a local hidden variable theory these non-local transition sets must be empty $T^{a\leftrightarrow a'}_{\ \ b}=T^{\ a}_{b\leftrightarrow b'}=T^{a\leftrightarrow a'}_{\ \ b'}=T^{\ a'}_{b\leftrightarrow b'}=\emptyset$.

One might ask how probable it is for a $\lambda$ to belong to the transition set $T^{\ a}_{b\leftrightarrow b'}$. In view of eq. (\ref{transdef}) the probability/measure is given by \cite{Valentini} 
\begin{eqnarray}\label{transprob}
P(T^{\ a}_{b\leftrightarrow b'})=\int_{\lambda\in T^{\ a}_{b\leftrightarrow
b'}}\rho(\lambda)d\lambda=\frac{1}{2}\int_\Lambda|A(a,b,\lambda)-A(a,b',\lambda)|\rho(\lambda)d\lambda          
\end{eqnarray}
where $\rho(\lambda)$ is a probability distribution of hidden variables.\footnote{Note that in eq. (\ref{transprob}) we are making use of the assumption that $\rho(\lambda)$ is independent of the angles.} Similarly the probability for a $\lambda$ to belong to $T^{a\leftrightarrow a'}_{\ \ b}$ is given by
\begin{eqnarray}
P(T^{a\leftrightarrow a'}_{\ \ b})=\int_{\lambda\in T^{a\leftrightarrow a'}_{\ \ b}}\rho(\lambda)d\lambda=\frac{1}{2}\int_\Lambda|B(a,b,\lambda)-B(a',b,\lambda)|\rho(\lambda)d\lambda.        
\end{eqnarray}
It is important realize that the probabilities $P(T^{\ a}_{b\leftrightarrow b'})$ and $P(T^{a\leftrightarrow a'}_{\ \ b})$ are critically dependent on the choice of the distribution of hidden variables $\rho(\lambda)$. More generally, if a hidden variable theory can reproduce the statistics of quantum mechanics then there exists a special `equilibrium' distribution $\rho_{eq.}(\lambda)$ so that one reproduces the quantum statistics for all possible experiments if and only if $\rho(\lambda)=\rho_{eq.}(\lambda)$ (see also \cite{Valentinithesis,Htheorem,Dynorigin,DGZ}). We shall assume the distribution $\rho(\lambda)$ to be the equilibrium one except in sections \ref{signalocality} and where we shall graphically illustrate the signal-locality theorem by Valentini \cite{Valentini,Valentinib}.
\section{Contextuality vs Non-Locality}\label{CvsNL}
%
Using the notion of transition sets, non-contextuality can be defined as follows:

\begin{quote} {\bf Non-Contextuality}. {\it Let $\{\hat{\mathcal{A}};\hat{\mathcal{B}},\hat{\mathcal{C}}\}$ be triplet of quantum observables satisfying the commutation relations $[\hat{\mathcal{A}},\hat{\mathcal{B}}]=[\hat{\mathcal{A}},\hat{\mathcal{C}}]=0$ and $[\hat{\mathcal{B}},\hat{\mathcal{C}}]\neq0$. Let $A_B(\lambda)$ refer to the outcome of the measurement of observable $\hat{\mathcal{A}}$ when measured together with $\hat{\mathcal{B}}$ and similarly for $A_C(\lambda)$. A transition set is then defined by $T^{\ \ \hat{\mathcal{A}}}_{\hat{\mathcal{B}}\leftrightarrow \hat{\mathcal{C}}}=\{\lambda\in\Lambda|A_B(\lambda)\neq A_C(\lambda)\}$. A hidden variable theory is then said to be non-contextual if, for all possible choices of observable triplets $\{\hat{\mathcal{A}};\hat{\mathcal{B}},\hat{\mathcal{C}}\}$, the corresponding transition set $T^{\ \ \hat{\mathcal{A}}}_{\hat{\mathcal{B}}\leftrightarrow \hat{\mathcal{C}}}$ is empty, i.e. $T^{\ \ \hat{\mathcal{A}}}_{\hat{\mathcal{B}}\leftrightarrow \hat{\mathcal{C}}}=\emptyset$}.
\end{quote}
Let us now see what non-contextuality amounts to in the EPRB gedanken experiment. As before consider the two spatially separated spacetime regions ${\sf A}$ and ${\sf B}$ each containing an apparatus `measuring' the spin. This means that we are dealing with the four observables $\hat{\mathcal{A}}={\bf a}\cdot\boldsymbol{\sigma}\otimes \boldsymbol{1}$, $\hat{\mathcal{A}'}={\bf a}'\cdot \boldsymbol{\sigma}\otimes \boldsymbol{1}$, $\hat{\mathcal{B}}={\bf 1}\otimes \boldsymbol{b}\cdot\boldsymbol{\sigma}$, and $\hat{\mathcal B}'={\bf 1}\otimes b'\cdot\boldsymbol{\sigma}$, where ${\bf \sigma}$ are the usual Pauli matrices.

For typical choices of angles these observables satisfy the following commutation relations:
\begin{eqnarray}
[\hat{\mathcal{A}},\hat{\mathcal{B}}]=[\hat{\mathcal{A}},\hat{\mathcal B}']=0\qquad [\hat{\mathcal{A}}',{\cal
B}]=[\hat{\mathcal{A}}',\hat{\mathcal B}']=0\qquad [\hat{\mathcal{A}},\hat{\mathcal{A}}']\neq0\neq [{\cal
B},\hat{\mathcal B}']. 
\end{eqnarray}
In order to make connection with the definition of non-contextuality we single out the following four observable triplets
\begin{eqnarray}
\{\hat{\mathcal{B}};\hat{\mathcal{A}},\hat{\mathcal{A}}'\}\qquad \{{\cal
B'};\hat{\mathcal{A}},\hat{\mathcal{A}}'\}\qquad \{\hat{\mathcal{A}};\hat{\mathcal{B}},\hat{\mathcal B}'\}\qquad\{\hat{\mathcal{A}}';{\cal
B},\hat{\mathcal B}'\}.
\end{eqnarray}
Non-Contextuality for the EPRB case simply means that, in particular, the transition sets are all empty, i.e. 
\begin{eqnarray}
T^{a\leftrightarrow a'}_{\ \ b}=T^{a\leftrightarrow a'}_{\ \
b'}=T_{b\leftrightarrow b'}^{\ a}=T_{b\leftrightarrow b'}^{\
a'}=\emptyset. 
\end{eqnarray}
Here we have used, as before, the angles $\{a,a',b,b'\}$ to denote a transition set rather than the observables $\{\hat{\mathcal{A}},\hat{\mathcal{A}}',\hat{\mathcal{B}},\hat{\mathcal B}'\}$. The emptiness of these sets immediately implies that $A(a,b,\lambda)=A(a,\lambda)$ and $B(a,b,\lambda)=B(a,\lambda)$. Thus, non-contextuality implies locality. (The converse is not true.) 

%
\section{The impossibility of local hidden variable theories}\label{logical}
%
In this section we will establish a direct logical contradiction between non-contextuality /locality and the statistics of quantum mechanics. We do that by first showing that a particular set is non-empty. Then using the non-emptiness of that set we show that assuming non-contextuality/locality yields a direct logical contradiction.
\begin{figure}
\begin{center}
\psfrag{a}{$S_1^-$}
\psfrag{b}{$S_2^-$}
\psfrag{c}{$S_3^-$}
\psfrag{d}{$S_4^+$}
\psfrag{Lambda space}{{\Large $\Lambda$-space}}
\epsfig{figure=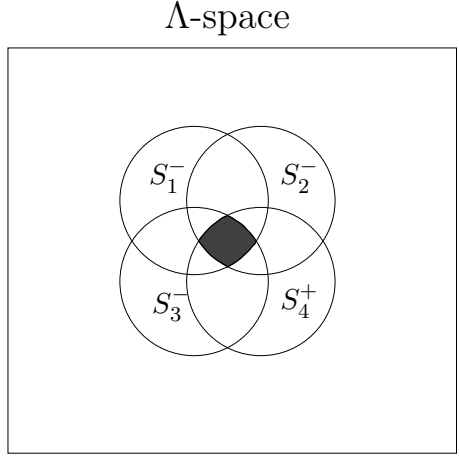,width=6cm}
\end{center}
\caption{The four circles correspond to regions in the hidden variable space for which the product $AB$ of the outcome takes the values $-1,-1,-1$, and $+1$ for the angular choices $\{a,b\}$, $\{a',b\}$, $\{a',b'\}$, and $\{a,b'\}$ respectively. The dark region is the intersection $S_{1}^-\cap S_{2}^-\cap S_{3}^-\cap S_{4}^+$ of all four sets.} 
\label{foursets}
\end{figure}
Let us consider\footnote{This approach is adapted after Hardy \cite{Hardy}.} a list of four statements:
$s_1^-,s_2^-,s_3^-,s_4^+$
\begin{eqnarray}\label{stat}
A(a,b,\lambda)B(a,b,\lambda)&=&-1\\
A(a',b,\lambda)B(a',b,\lambda)&=&-1\\
A(a',b',\lambda)B(a',b',\lambda)&=&-1\\
A(a,b',\lambda)B(a,b',\lambda)&=&+1.
\end{eqnarray}
The reason for choosing these particular signs will become clear in section \ref{relation}. We can now ask the question: In a hidden variable theory capable of reproducing the quantum statistics, how probable is it for all those four statements to be true at the same time? Let $S_1^-,S_2^-,S_3^-,S_4^+$ denote the sets in the hidden variable space $\Lambda$ so that the statement $s_1^-$,..., $s_4^+$ is true if and only if $\lambda\in S_1^-$,..., $\lambda\in S_4^+$ respectively. The question then becomes: what is the probability $P(S_{1}^-\cap S_{2}^-\cap S_{3}^-\cap S_{4}^+)$? See Fig. \ref{foursets} for a graphical illustration of this set.

A lower bound on $P(S_{1}^-\cap S_{2}^-\cap S_{3}^-\cap S_{4}^+)$ may be obtained \cite{Hardy} as follows ($S^c$ stands for the complement
of $S$): 
\begin{eqnarray}
1-P(S_{1}^-\cap S_{2}^-\cap S_{3}^-\cap S_{4}^+)&=&P([S_{1}^-\cap
S_{2}^-\cap S_{3}^-\cap S_{4}^+]^c)\nonumber\\&=&P([S_{1}^-]^c\cup
[S_{2}^-]^c\cup [S_{3}^-]^c\cup [S_{4}^+]^c)\nonumber\\&\leq&
P([S_1^-]^c)+P([S_2^-]^c)+P([S_3^-]^c)+P([S_4^+]^c)\nonumber\\&=&
4-P(S_1^-)+P(S_2^-)+P(S_3^-)+P(S_4^+)
\end{eqnarray}
which yields the lower bound

\begin{eqnarray}\label{lowb}
P(S_{1}^-\cap S_{2}^-\cap S_{3}^-\cap S_{4}^+)\geq
P(S_1^-)+P(S_2^-)+P(S_3^-)+P(S_4^+)-3.
\end{eqnarray}
We shall hereafter refer to lower bounds of this type as Hardy lower bounds. The lower bound (\ref{lowb}) can be computed for the singlet state using the familiar quantum statistics
\begin{eqnarray}\label{quantstat}
P(S_1^\pm)&=&P\left(\{\lambda\in\Lambda|A(a,b,\lambda)B(a,b,\lambda)=\pm1\right\})=\frac{1\mp\cos\theta_{ab}}{2}\nonumber\\
P(S_2^\pm)&=&P\left(\{\lambda\in\Lambda|A(a',b,\lambda)B(a',b,\lambda)=\pm1\right\})=\frac{1\mp\cos\theta_{a'b}}{2}\nonumber\\
P(S_3^\pm)&=&P\left(\{\lambda\in\Lambda|A(a',b',\lambda)B(a',b',\lambda)=\pm1\right\})=\frac{1\mp\cos\theta_{a'b'}}{2}\nonumber\\
P(S_4^\pm)&=&P\left(\{\lambda\in\Lambda|A(a,b',\lambda)B(a,b',\lambda)=\pm1\right\})=\frac{1\mp\cos\theta_{ab'}}{2}.
\end{eqnarray}
Inserting this particular statistics into eq. (\ref{lowb}) with the specific angles $a-b=b-a'=a'-b'=\theta$ and $a-b'=3\theta$ yields
\begin{eqnarray}
\!\!\!\!\!\!\!\!&&P(S_{1}^-\cap S_{2}^-\cap S_{3}^-\cap S_{4}^+)\nonumber\\&\geq&\frac{1}{2}(1+\cos\theta)+\frac{1}{2}(1+\cos\theta)+\frac{1}{2}(1+\cos\theta)+\frac{1}{2}(1-\cos3\theta)-3\nonumber\\&=&\frac{1}{2}(3\cos\theta-\cos 3\theta)-1 
\end{eqnarray}
which is equal to $\sqrt{2}-1$ when $\theta=\frac{\pi}{4}$, for example.

We have now seen that, in order to reproduce the quantum statistics, the set $S_{1}^-\cap S_{2}^-\cap S_{3}^-\cap S_{4}^+$ must for some choices of angles $\{a,a',b,b'\}$ have a measure greater than zero and is therefore nonempty. Now we have a clearcut logical contradiction with the non-contextuality assumption, $T^{a\leftrightarrow a'}_{\ \ b}=T^{a\leftrightarrow a'}_{\ \ b'}=T_{b\leftrightarrow b'}^{\ a}=T_{b\leftrightarrow b'}^{\ a'}=\emptyset$. To see this let $\lambda\in S_{1}^-\cap S_{2}^-\cap S_{3}^-\cap S_{4}^+$ and contemplate the following logical chain of deductions:
\begin{eqnarray}\label{contradict}
A(a,b,\lambda)=\pm 1\qquad&\stackrel{\lambda\in S_1^-}{\Rightarrow}&\qquad B(a,b,\lambda)=\mp
1\nonumber\\&\ &\qquad\qquad\Downarrow T^{a\leftrightarrow a'}_{\ \ b}=\emptyset\nonumber\\A(a',b,\lambda)=\pm
1\qquad&\stackrel{\lambda\in S_2^-}{\Leftarrow}&\qquad B(a',b,\lambda)=\mp
1\nonumber\\ T^{\ a'}_{b\leftrightarrow
b'}=\emptyset\Downarrow\qquad\qquad\nonumber\\A(a',b',\lambda)=\pm
1\qquad&\stackrel{\lambda\in S_3^-}{\Rightarrow}&\qquad
B(a',b',\lambda)=\mp 1\nonumber\\&\ 
&\qquad\qquad\Downarrow T^{a\leftrightarrow a'}_{\ \
b'}=\emptyset\nonumber\\A(a,b',\lambda)=\mp 1\qquad&\stackrel{\lambda\in S_4^+}{\Leftarrow}&\qquad
B(a,b',\lambda)=\mp 1\nonumber\\ T^{\
a}_{b\leftrightarrow
b'}=\emptyset\Downarrow\qquad\qquad\nonumber\\A(a,b,\lambda)=\mp
1\qquad&\Rightarrow&\qquad \perp. 
\end{eqnarray}
In short, if $A(a,b',\lambda)=\pm1$ then $A(a,b,\lambda)=\mp1$ which is
a clear logical contradiction. Therefore, no non-contextual/local hidden
variable theory can reproduce the quantum statistics. 

Is interesting to note the limited role probability plays in the above
proof. The only thing needed is the non-emptiness of the set $S_{1}^-\cap S_{2}^-\cap
S_{3}^-\cap S_{4}^+$ and that is established by showing that it has a
positive measure. Within our approach it is not of great interest whether one can obtain a logical contradiction for all $\lambda\in\Lambda$ and for any quantum state. Instead, a theory will be said to be {\em contextual} if {\em some} transition set is non-empty.
%
\section{Relation to Bell type inequalities}\label{relation}
%
The choice of signs $(-,-,-,+)$ for the four statements
$s_1^-,s_2^-,s_3^-,s_4^+$ in (\ref{stat}) is not the only combination
that gives rise to a contradiction with non-contextuality. All combinations 
where the product of the signs is negative will in the same way
contradict the non-contextuality assumption if the corresponding set
is nonempty. The choices of signs for which the product is
negative are:
\begin{eqnarray}
(-,+,+,+)\qquad (+,-,+,+)\qquad (+,+,-,+)\qquad (+,+,+,-)\nonumber\\
(+,-,-,-) \qquad(-,+,-,-) \qquad(-,-,+,-) \qquad(-,-,-,+).\nonumber
\end{eqnarray}
Note that these different choices of signs refer to disjoint sets in
$\Lambda$ since the outcomes are different. In the following it shall prove useful to work with the union of these
disjoint sets. Therefore, let $\sigma_-\subseteq\Lambda$ be the set
for which the product of the signs is negative. Introduce the
convenient short-hand notation: 
\begin{eqnarray}
p_{1}^\pm&=&P(S_1^\pm)\\
p_{2}^\pm&=&P(S_2^\pm)\\
p_{3}^\pm&=&P(S_3^\pm)\\
p_{4}^\pm&=&P(S_4^\pm).
\end{eqnarray}
In appendix \ref{appA} a single unified non-negative lower bound on $P(\sigma_-)$ is derived:
\begin{eqnarray}\label{lowbound}
P(\sigma_-)&\geq&\frac{1}{2}(|p_1^+-p_2^-|+|p_3^+-p_4^+|+|p_1^+-p_2^+|+|p_3^+-p_4^-|\nonumber\\&+&
\left||p_1^+-p_2^-|+|p_3^+-p_4^+|-1\right|+\left||p_1^+-p_2^+|+|p_3^+-p_4^-|-1\right|)-1.
\end{eqnarray}
A single unified Bell inequality
\begin{eqnarray}\label{Bellineq}
|p_1^+-p_2^-|+|p_3^+-p_4^+|+|p_1^+-p_2^+|+|p_3^+-p_4^-|\nonumber&&\nonumber\\+\left||p_1^+-p_2^-|+|p_3^+-p_4^+|-|p_1^+-p_2^+|-|p_3^+-p_4^-|\right|&\leq&2    
\end{eqnarray} 
is also derived that contains all other inequalities (that involves two
possible choices of angles at ${\sf A}$ and ${\sf B}$) as special
cases. It is also seen in appendix \ref{appA} that the lower bound of
$P(\sigma_-)$ is half of a any violation of that unified Bell
inequality. We will therefore simply refer to the right hand side of
eq. (\ref{lowbound}) as the `violations of Bell inequalities'.
%
\section{What do violations of Bell inequalities quantify?}\label{Quantify}
%
Using the notion of transition sets we now proceed to show how
one can graphically illustrate what violations of Bell inequalities
quantify. Recall the logical chain of deductions
(\ref{contradict}). In what way can we escape a contradiction? Clearly
we must abandon the non-contextuality assumption that requires all
transition sets to be empty. One way to avoid contradiction is the
following: 
\begin{eqnarray}\label{avoidcontr}
A(a,b,\lambda)=\pm 1\qquad&\stackrel{\lambda\in S_1^-}{\Rightarrow}&\qquad B(a,b,\lambda)=\mp
1\nonumber\\&\ &\qquad\qquad\Downarrow\lambda\notin T^{a\leftrightarrow a'}_{\ \ b}\nonumber\\A(a',b,\lambda)=\pm
1\qquad&\stackrel{\lambda\in S_2^-}{\Leftarrow}&\qquad B(a',b,\lambda)=\mp
1\nonumber\\\lambda\notin T^{\ a'}_{b\leftrightarrow
b'}\Downarrow\qquad\qquad\nonumber\\ A(a',b',\lambda)=\pm
1\qquad&\stackrel{\lambda\in S_3^-}{\Rightarrow}&\qquad B(a',b',\lambda)=\mp 1\nonumber\\&\
&\qquad\qquad\Downarrow \lambda\notin T^{a\leftrightarrow a'}_{\ \
b'}\nonumber\\A(a,b',\lambda)=\mp 1\qquad&\stackrel{\lambda\in S_4^+}{\Leftarrow}&\qquad
B(a,b',\lambda)=\mp 1\nonumber\\\lambda\in T^{\
a}_{b\leftrightarrow b'}\Downarrow\qquad\qquad\nonumber\\A(a,b,\lambda)=\pm 1\qquad
\end{eqnarray}
that is, $\lambda\in[T^{a\leftrightarrow a'}_{\ \ b}]^c\cap[T^{\
a'}_{b\leftrightarrow b'}]^c\cap[T^{a\leftrightarrow a'}_{\ \
b'}]^c\cap T^{\ a}_{b\leftrightarrow b'}$. In total there are eight
different ways of escaping the contradiction corresponding to
$\lambda$ being in precisely one of the four transition sets or
precisely three. That is, $\lambda$ is in one of the following eight disjoint sets:
\begin{eqnarray}\label{eightsets}
\lambda&\in&T_1=[T^{a\leftrightarrow a'}_{\ \ b}]^c\cap T^{\ a'}_{b\leftrightarrow b'}\cap T^{a\leftrightarrow a'}_{\ \ b'}\cap T^{\ a}_{b\leftrightarrow b'}\\
\lambda&\in&T_2= T^{a\leftrightarrow a'}_{\ \ b}\cap[T^{\ a'}_{b\leftrightarrow b'}]^c\cap T^{a\leftrightarrow a'}_{\ \ b'}\cap T^{\ a}_{b\leftrightarrow b'}\\
\lambda&\in&T_3= T^{a\leftrightarrow a'}_{\ \ b}\cap T^{\ a'}_{b\leftrightarrow b'}\cap[T^{a\leftrightarrow a'}_{\ \ b'}]^c\cap T^{\ a}_{b\leftrightarrow b'}\\
\lambda&\in&T_4= T^{a\leftrightarrow a'}_{\ \ b}\cap T^{\ a'}_{b\leftrightarrow b'}\cap T^{a\leftrightarrow a'}_{\ \ b'}\cap [T^{\ a}_{b\leftrightarrow b'}]^c\\
\lambda&\in&T_5=[T^{a\leftrightarrow a'}_{\ \ b}]^c\cap[T^{\ a'}_{b\leftrightarrow b'}]^c\cap[T^{a\leftrightarrow a'}_{\ \ b'}]^c\cap T^{\ a}_{b\leftrightarrow b'}\\
\lambda&\in&T_6=[T^{a\leftrightarrow a'}_{\ \ b}]^c\cap[T^{\ a'}_{b\leftrightarrow b'}]^c\cap T^{a\leftrightarrow a'}_{\ \ b'}\cap [T^{\ a}_{b\leftrightarrow b'}]^c\\\lambda&\in&T_7=[T^{a\leftrightarrow a'}_{\ \ b}]^c\cap T^{\ a'}_{b\leftrightarrow b'}\cap[T^{a\leftrightarrow a'}_{\ \ b'}]^c\cap [T^{\ a}_{b\leftrightarrow b'}]^c\\
\lambda&\in&T_8= T^{a\leftrightarrow a'}_{\ \ b}\cap[T^{\ a'}_{b\leftrightarrow b'}]^c\cap[T^{a\leftrightarrow a'}_{\ \ b'}]^c\cap [T^{\ a}_{b\leftrightarrow b'}]^c.\label{eightsetslast}
\end{eqnarray}
\begin{figure}
\begin{center}
\psfrag{Lambda space}{{\Large $\Lambda$-space}}
\psfrag{a}{$T^{a\leftrightarrow a'}_{\ \ b}$}
\psfrag{b}{$T^{\ a'}_{b\leftrightarrow b'}$}
\psfrag{c}{$T^{a\leftrightarrow a'}_{\ \ b'}$}
\psfrag{d}{$T^{\ a}_{b\leftrightarrow b'}$}
\epsfig{figure=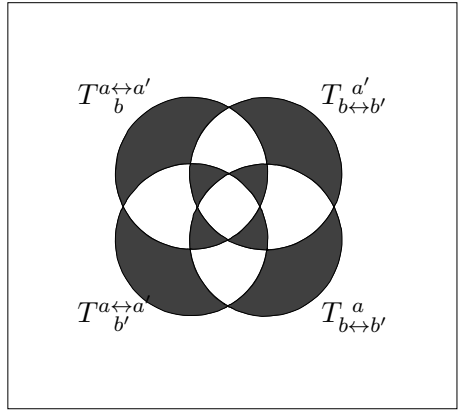,width=6cm}
\end{center}
\caption{The picture depicts the four (in general intersecting)
non-local transition sets 
$T^{a\leftrightarrow a'}_{\ \ b}$, $T^{\ a'}_{b\leftrightarrow
b'}$, $T^{a\leftrightarrow a'}_{\ \ b'}$, and $T^{\
a}_{b\leftrightarrow b'}$. Violations of Bell inequalities put a lower
bound only on the dark region. This is the region where $\lambda$
belongs to
precisely one or to precisely in three
non-local transition sets.}  
\label{transets} 
\end{figure}
In order to reproduce the statistics of quantum mechanics the measure
$P(\sigma_-)$ must be greater than or equal to the unified Hardy lower
bound (\ref{lowbound}). And whenever $\lambda\in\sigma_-$, then $\lambda$ will be in one of
the above eight sets. Therefore violations of Bell inequalities
provide a measure of the union of 
these eight (trivially) disjoint sets. This is graphically illustrated
in Fig. \ref{transets}.\footnote{For esthetic reasons the picture does
not display regions where $\lambda$ belongs precisely to
$T^{a\leftrightarrow a'}_{\ \ b}\cap[T^{\ a'}_{b\leftrightarrow
b'}]^c\cap[T^{a\leftrightarrow a'}_{\ \ b'}]^c\cap T^{\
a}_{b\leftrightarrow b'}$ or $[T^{a\leftrightarrow a'}_{\ \ b}]^c\cap
T^{\ a'}_{b\leftrightarrow b'}\cap T^{a\leftrightarrow a'}_{\ \
b'}\cap[T^{\ a}_{b\leftrightarrow b'}]^c$. These sets are disjoint from $\sigma_-$ and hence not so interesting to display.} The upper left, upper right,
lower left, and lower right 
circle represent respectively the four intersecting transition sets
$T^{a\leftrightarrow a'}_{\ \ b},T^{\ a'}_{b\leftrightarrow
b'},T^{a\leftrightarrow a'}_{\ \ b'}$ and $T^{\ a}_{b\leftrightarrow
b'}$. Violations of Bell inequalities provide a lower bound only on
the measure of the dark region. This set is where $\lambda$ either
belongs to precisely one transition set or precisely three.

 Note that the violations of Bell inequalities are unable to tell us anything about the measure of the regions where $\lambda$ is in precisely two non-local transition sets or
precisely four. In particular, this means that a hidden variable that reproduces the quantum statistics for a {\em fixed set} of possible angles $\{a,a',b,b'\}$ can be non-local without violating the Bell inequality (\ref{Bellineq}). This does not contradict a theorem due to Fine \cite{Fine82} who showed that if Bell the inequality (\ref{Bellineq}) is satisfied there {\em exists} a local hidden variable theory that reproduces the statistics. The theorem does not state that all hidden variable theories that satisfies the specific Bell inequality (\ref{Bellineq}) are local.
%
\section{Communication cost}\label{communication}
%
It is interesting to see how much communication is needed between the spatially separated regions ${\sf A}$ and ${\sf B}$ in order to reproduce the quantum correlations for a fixed set of possible angles \{$a,a',b,b'$\}. Here we will reproduce a result of Pironio \cite{Pironio} by a simple graphical inspection of Fig. \ref{transets}.

Suppose Alice and Bob agree to play the following game. At a point C they are both provided with the same random parameter $\lambda$. Then Alice and Bob walk off to the regions ${\sf A}$ and ${\sf B}$ respectively. At their respective locations they are each provided with an angle, $a$ or $a'$ for Alice and $b$ or $b'$ for Bob. It is assumed that the angles at ${\sf A}$ and ${\sf B}$ are randomly picked with a 50-50\% probability.

Alice and Bob should each present an `outcome', $+1$ or $-1$. Their task is to come up with strategy (i.e. a way to assign outcomes given the random parameter $\lambda$ and angles) so that in the long run, playing the game several times, they reproduce the quantum statistics (\ref{quantstat}). This they might do with aid of their random parameter received in region C and the angles they are assigned at ${\sf A}$ and ${\sf B}$. 

If Alice and Bob are not allowed to communicate the angles they were assigned, the best they can do is to make up a strategy that makes use of the common random parameter $\lambda$ and their respective angle, i.e. strategies of the type $A(a,\lambda)$ and $B(b,\lambda)$. In this sense the assignment of outcomes is non-contextual since an individual outcome will not depend on what angle the other person was assigned. But as we know, no such a strategy can reproduce the statistics of quantum mechanics and we will end up with the logical contradiction (\ref{contradict})

However, there is another class of strategies that will work. If the random variable $\lambda$ happens to belong to some particular subset of $\Lambda$ then the strategy could require Alice and/or Bob to make use of the angle that was assigned to the other. In this case the angles has to be communicated from {\sf A} to {\sf B} or {\sf B} to {\sf A}. These strategies will be of the type $A(a,b,\lambda)$ and $B(a,b,\lambda)$ and the subsets of $\Lambda$ for which the outcome at ${\sf A}/{\sf B}$ depends on the choice of angle at ${\sf B}/{\sf A}$ are the transition sets $T^{a\leftrightarrow a'}_{\ \ b}$, $T^{a\leftrightarrow a'}_{\ \ b'}$, $T_{b\leftrightarrow b'}^{\ a}$, and $T_{b\leftrightarrow b'}^{\ a'}$.

Suppose then that the shared random parameter $\lambda$ happens to belong to, say, $T_8$. This is the region for which $\lambda$ lies in the transition set $T^{a\leftrightarrow a'}_{\ \ b}$ and in no other transition set. If Bob was assigned the angle $b$ then Bob faces the dilemma that in order to carry out the strategy Bob must know about Alice's angle. Therefore Bob must receive information about the angle Alice was assigned before he can declare his outcome. Since for Alice there is only two angular settings, one bit of information will suffice to communicate the angle to Bob. However, if Bob was assigned the angle $b'$ his outcome is not dependent on Alice's angle (the sets $T_8$ and $T^{a\leftrightarrow a'}_{\ \ b'}$ are disjoint.). Nevertheless, there is no way for Alice to know what angle Bob got assigned and for `safety' she must communicate her angle to Bob even in this case. Thus, 1 bit of information must be sent whenever $\lambda$ happens to belong to $T_8$.\footnote{One could also contemplate other communication strategies. For example, Bob might send a one-bit signal to Alice whenever he happens to need information about her angle. So whenever Bob was assigned the angle $b$ and $\lambda$ was in the set $T_8$ two bits of information must be sent: one bit Bob telling Alice he needs information and one bit when Alice reveals the angle she was assigned. But in 50\% of the cases Bob will be assigned the angle $b'$ in which case 0 bits is sent. The average number of bits will then still be 1.}

Consider now the possibility that the parameter $\lambda$ belongs to $T_1$, that is, $\lambda$ belongs to the three transition sets $T^{\ a'}_{b\leftrightarrow b'}$, $T^{a\leftrightarrow a'}_{\ \ b'}$ and  $T^{\ a}_{b\leftrightarrow b'}$ but not to $T^{a\leftrightarrow a'}_{\ \ b}$. Then no matter what angle Alice gets assigned she is going to need information about the angle Bob has been assigned. However, only when Bob gets assigned the angle $b'$ he is going to need information about Alice angle, but again Alice cannot know that and she has to send one bit even in that case. Thus, 2 bits of information must be sent whenever $\lambda$ happens to belong to $T_1$. 

What about the regions in Fig. \ref{transets} where $\lambda$ belongs to precisely two or four non-local transition sets? If $\lambda$ belongs to one of these regions communication is required. Interestingly, the unified lower bound eq. (\ref{lowbound}) does not reveal whether these regions have zero measure or not. Therefore Alice and Bob can come up with strategies that involve non-local communication  without violating the Bell inequality (\ref{Bellineq}). For example, if the parameter $\lambda$ belongs to $T^{\ a'}_{b\leftrightarrow b'}$ and $T^{a\leftrightarrow a'}_{\ \ b'}$, but not to $T^{\ a}_{b\leftrightarrow b'}$ and $T^{a\leftrightarrow a'}_{\ \ b}$, then 2 bits needs to be sent. However, if $\lambda$ belongs to, for example, $T^{\ a'}_{b\leftrightarrow b'}$ and $T^{\ a}_{b\leftrightarrow b'}$, but not to $T^{a\leftrightarrow a'}_{\ \ b'}$ and $T^{a\leftrightarrow a'}_{\ \ b}$, then only 1 bit needs to be sent from Bob to Alice.

Nevertheless, the quantum statistics does not require these regions to have nonzero measure. Therefore we can neglect those regions in our analysis which was aimed at deriving a {\em lower bound} on communication cost. Thus, putting the measure of these regions to zero, the average bits of information per run needed is
\begin{eqnarray}
\bar{b}&=&2\cdot(P(T_1)+ P(T_2)+P(T_3)+P(T_4))\\
&+&1\cdot(P(T_5)+P(T_6)+P(T_7)+P(T_8)).
\end{eqnarray}
Since quantum mechanics only reveals the sum $\sum_i P(T_i)=P(\sigma_-)$ it is clear from the above expression that one cannot compute $\bar{b}$ without fixing the sum $\sum_{i=1}^4 P(T_i)$. However, by convexity $\bar{b}$ has a minimum value. The minimum value equal to $P(\sigma_-)$ is obtained when $P(T_1)=P(T_2)=P(T_3)=P(T_4)=0$. Thus we have $\bar{b}\geq P(\sigma_-)$ which is the result of \cite{Pironio}.

We end this section by noting that Maudlin \cite{Maudlin} has shown that if one choose angles $a,b$ from the set $[0,\pi]$ \footnote{Maudlin analysis involves photons rather than electrons so that the range of angles is $[0,\pi]$ rather than $[0,2\pi]$.} with uniform probability then in average a minimum of $1.174$ bits per run must be sent between Alice and Bob. However, Maudlin makes use of a specific scheme and therefore it is not entirely clear whether or not this communication cost can be reduced further. 

\section{Signal locality and non-contextuality of statistics}\label{signalocality}
Even though the the individual outcomes in the EPRB scenario has to depend on the choice of the non-local angle the statistics is remarkable independent on such a choice. As was pointed out by Valentini \cite{Valsignlocal,PearlenValentini,Valentinithesis}, this non-contextuality of the statistics is a feature of the quantum equilibrium distribution $\rho_{eq}(\lambda)$ and for a different distribution $\rho(\lambda)\neq\rho_{eq}(\lambda)$ the statistics becomes contextual as well. We explore this now in more detail.

Following Valentini \cite{Valentini,Valentinib,PearlenValentini}, the transition set $T^{a\leftrightarrow a'}_{\ \ b}$ may be partitioned into two disjoint subsets 
\begin{eqnarray}
T^{a\leftrightarrow a'}_{\ \ b}(+,-)=\{\lambda\in\Lambda|B(a,b,\lambda)=+1,B(a',b,\lambda)=-1\}\label{trans1}\\
T^{a\leftrightarrow a'}_{\ \ b}(-,+)=\{\lambda\in\Lambda|B(a,b,\lambda)=-1,B(a',b,\lambda)=+1\}\label{trans2}
\end{eqnarray}
For the equilibrium distribution $\rho_{eq.}(\lambda)$ the probability for getting $B=+1$ is independent of the choice of angle at {\sf A}. This immediately implies that the measures of the sets must be equal equal $P(T^{a\leftrightarrow a'}_{\ \ b}(+,-))=P(T^{a\leftrightarrow a'}_{\ \ b}(-,+))$ where
\begin{eqnarray}
P(T^{a\leftrightarrow a'}_{\ \ b}(+,-))=\int_{\lambda\in T^{a\leftrightarrow a'}_{\ \ b}(+,-)}\rho_{eq.}(\lambda)d\lambda\nonumber\\P(T^{a\leftrightarrow a'}_{\ \ b}(-,+))=\int_{\lambda\in T^{a\leftrightarrow a'}_{\ \ b}(-,+)}\rho_{eq.}(\lambda)d\lambda\nonumber.
\end{eqnarray}
However, for a non-equilibrium distribution $\rho(\lambda)\neq\rho_{eq.}(\lambda)$ one might very well have $P(T^{a\leftrightarrow a'}_{\ \ b}(+,-))>0$ and  $P(T^{a\leftrightarrow a'}_{\ \ b}(-,+))=0$, for example. That implies conversely that the marginal statistics at ${\sf B}$ is sensitive to what measurement ($a$ or $a'$) is done at ${\sf A}$. Thus it is possible to transmit signals faster than the speed of light. Clearly, the result is quite general and holds for a large class of hidden variable theories including deBroglie-Bohm theory \cite{Valsignlocal}. However, as we have pointed out in section \ref{Trans}, there are non-trivial implicit assumptions made about the form of the outcome functions $A$ and $B$.

\begin{figure}
\begin{center}
\psfrag{B}{{\Huge B}}
\psfrag{+}{$+$}
\psfrag{-}{$-$}
\epsfig{figure=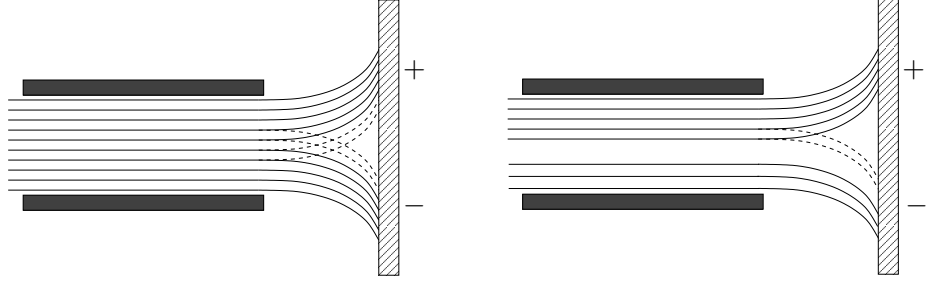,width=12cm}
\end{center}
\caption{The pictures illustrate a Stern-Gerlach apparatus in region $\sf B$. The apparatus in region $\sf A$ (not in the picture) can be aligned either along $a$ or $a'$. The solid lines refer to the outcome at ${\sf B}$ had the angle $a$ been chosen and the dashed if instead $a'$ had been chosen. The left picture illustrate the insensitivity of the marginal statistics in region ${\sf B}$ to the choice of angle at {\sf A} that occurs for the quantum equilibrium distribution. The right picture illustrates that the marginal statistics at $\sf B$ is sensitive to changes of angle at $\sf A$ when the distribution is not the quantum equilibrium one.}
\label{signloc} 
\end{figure}
Fig. \ref{signloc} illustrates the Stern-Gerlach apparatus in region ${\sf B}$. The solid lines refer to the outcome if angle $a$ is chosen at the distant Stern-Gerlach apparatus in region ${\sf A}$ (not in the picture) and the dashed line if instead $a'$ had been chosen. The left picture illustrates the detailed balancing that occurs for the quantum equilibrium distribution $\rho_{eq.}$. Because of the symmetric distribution $P(T^{a\leftrightarrow a'}_{\ \ b}(+,-))=P(T^{a\leftrightarrow a'}_{\ \ b}(-,+))$ there is no signal at the statistical level, only a `swap'. The right picture depicts a non-equilibrium ensemble with $P(T^{a\leftrightarrow a'}_{\ \ b}(+,-))>0$ and $P(T^{a\leftrightarrow a'}_{\ \ b}(-,+))=0$ for which the marginal statistics at ${\sf B}$ depends on the choice of angle at the spatially separated region ${\sf A}$.

%
\section{Measurement ordering contextuality}\label{MOC}
%
Any hidden variable theory capable of reproducing the quantum statistics must be contextual, i.e. there exists an observable triplet $\{\hat{\mathcal{A}};\hat{\mathcal{B}},\hat{\mathcal{C}}\}$, satisfying the commutation relations $[\hat{\mathcal{A}},\hat{\mathcal{B}}]=[\hat{\mathcal{A}},\hat{\mathcal{C}}]=0$ and $[\hat{\mathcal{B}},\hat{\mathcal{C}}]\neq0$, such that the corresponding transition set is non-empty $T^{\ \ \hat{\mathcal{A}}}_{\hat{\mathcal{B}}\leftrightarrow \hat{\mathcal{C}}}\neq\emptyset$. In this section we shall introduce a new form of contextuality: {\em measurement ordering contextuality}.

Consider the scenario where we first measure the operator $\hat{\mathcal{A}}$ and immediately record the outcome. Then after an arbitrary long time we decide to either measure $\hat{\mathcal{B}}$ or $\hat{\mathcal{C}}$, perhaps by flipping a coin. The hidden variable theory has to produce an outcome for $\hat{\mathcal{A}}$ before we have specified the context $\hat{\mathcal{B}}$ or $\hat{\mathcal{C}}$. The decision of measuring $\hat{\mathcal{B}}$ or $\hat{\mathcal{C}}$ might reasonably be regarded as a `free variable' \cite{Cuisine,Shimony}. Thus, unless we are willing to consider conspiratorial or retro-causal scenarios, the outcome for the first measurement has to be independent on context $\hat{\mathcal{B}}$ or $\hat{\mathcal{C}}$.\footnote{This is, of course, also the case in the deBroglie-Bohm hidden variable theory.} In contrast, if we measure $\hat{\mathcal{B}}$ or $\hat{\mathcal{C}}$ before measuring $\hat{\mathcal{A}}$ the outcome for $\hat{\mathcal{A}}$ could very well depend on the context. We therefore see that the transition set $T^{\ a}_{b\leftrightarrow b'}$ could be empty or non-empty depending on measurement order.

Note that we have assumed that the concept of `time ordering' for measurements is operationally well-defined. In special relativity this is not the case for space-like separated measurements. In the following we shall therefore confine the discussion to measurements which are time-like separated. 

We now turn to the proof of {\em measurement ordering contextuality} for general hidden variable theories. Let $_BA(\lambda)$ and $A_B(\lambda)$ respectively be the outcomes for the operator $\hat{\mathcal{A}}$ when $\hat{\mathcal{B}}$ is measured before and after the outcome of $\hat{\mathcal{A}}$ has been recorded. We can now define the following transition set:
\begin{eqnarray}\label{moctrans}
T_{\hat{\mathcal{A}}_{\hat{\mathcal{B}}}\leftrightarrow_{\hat{\mathcal{B}}}\hat{\mathcal{A}}}=\{\lambda\in\Lambda|_BA(\lambda)\neq A_B(\lambda)\}
\end{eqnarray}
A hidden variable theory is now said to be measurement ordering non-contextual if this transition set (\ref{moctrans}) is empty for all choices of commuting operator pairs $(\hat{\mathcal{A}},\hat{\mathcal{B}})$. This means in particular that $A_B(\lambda)=_B\!\!A(\lambda)$ and $A_C(\lambda)=_C\!\!A(\lambda)$ for all $\lambda\in\Lambda$. Assuming that the choice of measuring either $\hat{\mathcal{B}}$ or $\hat{\mathcal{C}}$ is a free variable we also have $A_B(\lambda)=A_C(\lambda)$ for all $\lambda\in\Lambda$. We can now establish the following chain of logical deductions:
\begin{eqnarray}
_BA(\lambda)=A_B(\lambda)=A_C(\lambda)=_C\!\!A(\lambda)
\end{eqnarray}
Thus, we can conclude that $_BA(\lambda)=_C\!\!A(\lambda)$ for all $\lambda\in\Lambda$ and for all operator triplets $\{\hat{\mathcal{A}};\hat{\mathcal{B}},\hat{\mathcal{C}}\}$ satisfying satisfying the specified commutation relations. The theory is therefore non-contextual according to the definition in section \ref{CvsNL}. Since no non-contextual theory can reproduce the statistics of quantum theory we have to give up measurement ordering non-contextuality. Hence, there exists a pair of commuting operators $\hat{\mathcal{A}}$ and $\hat{\mathcal{B}}$ such that the transition set $T_{\hat{\mathcal{A}}_{\hat{\mathcal{B}}}\leftrightarrow_{\hat{\mathcal{B}}}\hat{\mathcal{A}}}$ is non-empty.\footnote{Since the deBroglie-Bohm theory reproduces the quantum statistics and is neither retro-causal nor conspiratorial, it has to be measurement ordering contextual. In fact, this is the case as can be verified by explicit calculations for simple models.}

\section{Summary and outlook}
%
By using the notion of transition sets we provided a clear graphical illustration of what violations of Bell inequalities quantify (fig. \ref{transets}) as well as a direct logical contradiction between non-contextuality/locality in the EPRB gedanken experiment (\ref{contradict}). We also introduced a new form of quantum contextuality, measurement ordering contextuality, and showed that (excluding retro-causal or conspiratorial theories) is a feature of any hidden variable theory. This generalizes yet another feature of deBroglie-Bohm theory. Interestingly the hidden variable theory due to van Fraassen \cite{Fraassen,Redhead,Carsten} where one assigns values only to maximal (i.e. non-degenerate) Hermitian operators\footnote{Equivalently, one can view van Fraassen's model as assigning values to all projection valued measures, or equivalently to all bases of the Hilbert space in question.} seems to be ruled out since it is not measurement ordering contextual. More precisely, either it is a conspiratorial and/or a retro-causal model or it cannot reproduce the quantum statistics. 

As we have seen in this article the transition sets can be used to numerically quantify non-locality. Interestingly, the deBroglie-Bohm theory with von Neumann impulse measurements is much more non-local than required by the quantum statistics. In this sense the theory is too non-local. This will be the subject of a forthcoming paper.

\vspace{0.5cm}

\noindent{\large \bf Acknowledgments}\newline
This work begun as a collaboration with Antony Valentini to derive lower bounds on non-local information flow using methods in \cite{Hardy}. I want to thank Lucien Hardy for pointing out the reference \cite{Hardy} and Jonathan Barrett for pointing out reference \cite{Pironio}. I would also like to thank Antony Valentini for initial discussions and support, and Ward Struyve, Sebastiano Sonego, Owen Maroney, and Rob Spekkens for discussions and useful comments.\\\\
\appendix
\section{All Hardy bounds}\label{appA}
There are many combinations of signs that would lead to the
contradiction (\ref{contradict}). For each choice for which the
product is negative (eight different ways) there will be a corresponding Hardy lower bound:
\begin{eqnarray}
\alpha_1=p_1^++p_2^++p_3^++p_4^--3&\qquad&
\beta_1=p_1^-+p_2^-+p_3^-+p_4^+-3\label{HLB1}\\
\alpha_2=p_1^++p_2^++p_3^-+p_4^+-3&\qquad&
\beta_2=p_1^-+p_2^-+p_3^++p_4^--3\\
\alpha_3=p_1^++p_2^-+p_3^++p_4^+-3&\qquad&
\beta_3=p_1^-+p_2^++p_3^-+p_4^--3\\
\alpha_4=p_1^-+p_2^++p_3^++p_4^+-3&\qquad&
\beta_4=p_1^++p_2^-+p_3^-+p_4^--3.\label{HLB2}
\end{eqnarray}
In appendix \ref{appB} it is proved that at most one of the Hardy lower bounds (\ref{HLB1})--(\ref{HLB2}) can be greater than zero. It is therefore useful to compute $\max\{\alpha_1,...,\alpha_4,\beta_1,...,\beta_4\}$. This is readily done using the equality $\max\{a,b\}=\frac{1}{2}(a+b+|a-b|)$. In fact, the following formulas may easily be verified 

\begin{eqnarray}
\max\{\alpha_1,\alpha_2\}&=&p_1^++p_2^++|p_3^+-p_4^+|-2\\
\max\{\beta_1,\beta_2\}&=&p_1^-+p_2^-+|p_3^+-p_4^+|-2\\
\max\{\alpha_1,\alpha_2,\beta_1,\beta_2\}&=&|p_1^+-p_2^-|+|p_3^+-p_4^+|-1\label{Bellineq1}\\ 
\max\{\alpha_3,\alpha_4\}&=&p_3^++p_4^++|p_1^+-p_2^+|-2\\
\max\{\beta_3,\beta_4\}&=&p_3^-+p_4^-+|p_1^+-p_2^+|-2\\
\max\{\alpha_3,\alpha_4,\beta_3,\beta_4\}&=&|p_3^+-p_4^-|+|p_1^+-p_2^+|-1.\label{Bellineq2}
\end{eqnarray}
If we compute the max of (\ref{Bellineq1}) and (\ref{Bellineq2}) we end up with 
\begin{eqnarray}
\max\{\alpha_1,\alpha_2,...\beta_1,\beta_2...\}=\frac{1}{2}(|p_1^+-p_2^-|+|p_3^+-p_4^+|+|p_1^+-p_2^+|+|p_3^+-p_4^-|\nonumber\\+||p_1^+-p_2^-|+|p_3^+-p_4^+|-|p_1^+-p_2^+|+|p_3^+-p_4^-||)-1    
\end{eqnarray}
which can be both positive and negative for appropriate choices of angles. Requiring locality $\max\{\alpha_1,\alpha_2,...\beta_1,\beta_2...\}\leq 0$, yields a single unified Bell inequality\footnote{Using the definition of the correlation function $c(a,b)=p^+-p^-$, eq. (\ref{Bellineq1}) and (\ref{Bellineq2}) may readily be turned into the usual two Bell inequalities: $|c(a,b)\pm c(a,b')|+|c(a',b)\mp c(a',b')|\leq 2$. Since, for any statistical theory, at most one of the $\alpha$'s and $\beta$'s can be greater than zero, at most one of the two Bell inequalities can be violated. The single unified Bell inequality is violated only of some of the two usual inequalities is violated.} 
\begin{eqnarray}\label{singBell}
|p_1^+-p_2^-|+|p_3^+-p_4^+|+|p_1^+-p_2^+|+|p_3^+-p_4^-|\nonumber&+&\nonumber\\\left||p_1^+-p_2^-|+|p_3^+-p_4^+|-|p_1^+-p_2^+|-|p_3^+-p_4^-|\right|&\leq&2.  
\end{eqnarray} 
However, we shall be interested in obtaining a single unified lower bound which quantifies the violation of the unified Bell inequality (\ref{singBell}). Since probabilities are always positive we can replace the lower bounds (\ref{Bellineq1}) and (\ref{Bellineq2}) with  
\begin{eqnarray}
\frac{1}{2}(|p_1^+-p_2^-|+|p_3^+-p_4^+|-1+\left||p_1^+-p_2^-|+|p_3^+-p_4^+|-1\right|) 
\end{eqnarray}
and
\begin{eqnarray}
\frac{1}{2}(|p_3^+-p_4^-|+|p_1^+-p_2^+|-1+\left||p_3^+-p_4^-|+|p_1^+-p_2^+|-1\right|) 
\end{eqnarray}
respectively. Since these two expressions are always positive and refer to disjoint regions in the hidden variable space $\Lambda$ (outcomes are different) we can add them. This yields the single unified non-negative Hardy lower bound: 
\begin{eqnarray}
P(\sigma_-)&\geq&\frac{1}{2}(|p_1^+-p_2^-|+|p_3^+-p_4^+|+|p_1^+-p_2^+|+|p_3^+-p_4^-|\\&+& 
\left||p_1^+-p_2^-|+|p_3^+-p_4^+|-1\right|+\left||p_1^+-p_2^+|+|p_3^+-p_4^-|-1\right|)-1.
\end{eqnarray}
By the construction of $P(\sigma_-)$ it is easily seen that the lower bound on $P(\sigma_-)$ is half the amount of the violation of the unified Bell inequality (\ref{singBell}).

%
\section{Lemma}\label{appB}
{\it {\bf Lemma}: For any statistical theory at most one of the eight Hardy lower bounds can be greater than zero.} 
\newline\newline
{\bf Proof}:\newline
First note that since $p_i^\pm+p_i^\mp=1$ the sum of $\alpha_i$ and $\beta_i$ always equals $-2$. So, clearly, if $\alpha_i\geq0$ then $\beta_i\leq-2$.  

Next we show that if $\alpha_i\geq0$ then $\alpha_j,\beta_j\leq0$ for ($i\neq j$). Since the proof of this claim is identical for all combinations of $i,j$ it suffices to show it only for the case $i=1,j=2$. Since $\alpha_1=p_1^++p_2^++p_3^++p_4^--3\geq 0$ and probabilities are never greater than one we have $p_3^++p_4^-\geq 1$, or $1-p_3^+-p_4^-\leq 0$. Thus, 
\begin{eqnarray}
\alpha_2=p_1^++p_2^++p_3^-+p_4^+-3=\underbrace{p_1^++p_2^+}_{\leq
2}+(\underbrace{1-p_3^+-p_4^-}_{\leq 0})-2\leq 0.
\end{eqnarray} 
Also, because
\begin{eqnarray}
\alpha_2=p_1^++p_2^++p_3^-+p_4^+-3=\underbrace{p_1^++p_2^+}_{\geq   
1}+(\underbrace{1-p_3^+-p_4^-}_{\geq -1})-2\geq -2
\end{eqnarray}
and $\alpha_j+\beta_j=-2$ then $\beta_2\leq 0$. This concludes our proof.


\end{document}